DUST PARTICLE SIZE AND OPTICAL DEPTH ON MARS RETRIEVED BY THE MSL NAVIGATION CAMERAS

H. Chen-Chen*, S. Pérez-Hoyos, and A. Sánchez-Lavega.

Departamento de Física Aplicada I, Escuela de Ingeniería de Bilbao, Universidad del País Vasco (UPV/EHU). Bilbao 48013, Spain

* To whom correspondence should be addressed at: **hao.chen@ehu.eus**

**Abstract**
In this paper we show that Sun-viewing images obtained by the Mars Science Laboratory (MSL) Navigation Cameras (Navcam) can be used for retrieving the dust optical depth and constrain the aerosol physical properties at Gale Crater by evaluating the sky brightness as a function of the scattering angle. We have used 65 Sun-pointing images covering a period of almost three Martian years, from MSL mission sol 21 to sol 1646 (MY 31 to 33). Radiometric calibration and geometric reduction were performed on MSL Navcam raw image data records to provide the observed sky radiance as a function of the scattering angle for the near-Sun region (scattering angle from 4º to 30º). These curves were fitted with a multiple scattering radiative transfer model for a plane-parallel Martian atmosphere model using the discrete ordinates method. Modelled sky brightness curves were generated as a function of two parameters: the aerosol particle size distribution effective radius and the dust column optical depth at the surface. A retrieval scheme was implemented for deriving the parameters that generated the best fitting curve under a least-square error criterion. The obtained results present a good agreement with previous work, showing the seasonal dependence of both dust column optical depth and the effective particle radius.

**1. Introduction**

The dust aerosol suspended in the Martian atmosphere plays a key role in its climate, as the atmospheric thermal and dynamic structure, including the transport of the own aerosols, are mainly governed by the dust seasonal and spatial distribution and its radiative properties (e.g., Gierasch and Goody, 1972; Madeleine et al., 2011; Montabone et al. 2015). The dust radiative properties, in particular the single scattering albedo, extinction efficiency and phase function result from the dust refractive indices, particle shape and the size distribution of the particles (Pollack et al., 1995; Liou, 2002).

Many improvements have been achieved in our knowledge of Mars' airborne dust particle characteristics thanks to different exploration missions using both ground-based and orbital observations (e.g., Smith, 2008; Khare et al., 2017). In the case of remote sensing instruments on-board orbiting spacecraft, while they can provide a wider spatial and temporal data coverage, uncertainties arise in the retrieval process due to the similarity in composition between atmospheric dust and surface (Lemmon et al., 2015).

Studies of dust physical properties and its atmospheric loading from Mars' surface have been performed at several locations and periods with different instrumentation and techniques: the Viking landers' atmospheric imaging retrievals were used to evaluate the dust particle size distribution, single scattering albedo and phase function (Pollack et al., 1995). These properties were also derived using the Imager for Mars Pathfinder, which obtained Sun images and captured Martian sky brightness distribution at several wavelengths within the 440 to 965 nm range and characterised the atmospheric opacity (Smith and Lemmon, 1999) and dust properties (Tomasko et al., 1999; Markiewicz et al., 1999). The Mars Exploration Rovers (MER) mission also contributed to this subject with regular direct Sun imaging sequences using the solar filters of its Pancam imager (Lemmon et al., 2004) and retrieved the atmospheric opacity for a period of 5 Martian Years (Lemmon et al., 2015). At the Arctic region of Mars, the Phoenix Lander used a lidar instrument to obtain measurements of atmospheric dust loading and particle size distribution for 5 months (Komguem et al., 2013). Currently, the Mars Science Laboratory (MSL) mission has been characterising dust optical depth and aerosol size at its equatorial location using direct Sun imaging, passive sky spectroscopy and ultraviolet sensor systematic measurements (Lemmon et al., 2014; Smith et al., 2016, Vicente-Retortillo et al., 2017; McConnochie et al., 2017).

In this work we contribute to the study of Martian atmospheric dust particle physical properties by complementing those previous studies with independent retrievals of the particle effective radius and column optical depth and its temporal variations over Gale Crater using the data obtained from the MSL rover Curiosity navigation cameras (Navcam). We extend the temporal coverage of previous work until MSL sol 1646 and we validate the use of MSL Navcam observations for performing these retrievals.

Although not designed as a scientific instrument for atmospheric studies, images taken by rover's engineering cameras can be used as an alternative source of data for studying the atmospheric dust loading and deriving the aerosol physical properties. This was done, for instance, for the MER mission, by Soderblom et al., (2008a, 2008b), Smith and Wolff (2014), Wolfe and Lemmon (2015); and using MSL Curiosity Navcams by Moores et al., (2015) and Moore et al., (2016).

The capability of these cameras to obtain Mars' sky images under multiple geometry configurations, including observations very close to the Sun, allows the retrieval of the sky brightness as a function of the angle away from the solar disc centre (scattering angle), which can be evaluated to constrain dust aerosol particle size distribution and its shape (Tomasko et al., 1999). In particular, the sky brightness under a forward scattering scenario (up to 30º away from the Sun), is not sensitive to the aerosol optical properties (the refractive indices, i.e., composition) and shape; as for small scattering angles the intensity is dominated by the aerosol single scattering phase function and differences are negligible for spherical and non-spherical particles (Pollack et al., 1995; Liou, 2002).

We present a methodology for measuring the dust particle size distribution and retrieving its optical depth using MSL Navcam Sun-pointing images. In Section 2 we describe the Navcam observations dataset and the processing of the images used in this work. In Section 3 the methodology used to retrieve the dust aerosol optical depth and particle size distribution is presented. In Section 4 the results are shown, discussed and validated with retrievals from other instruments; and in Section 5 we provide a summary of the findings of this study and some future prospects.

**2. MSL Navcam observations**

The MSL rover is equipped with a set of 12 engineering cameras: 8 Hazard-Avoidance Cameras (Hazcams) and 4 Navigation Cameras (Navcams). The objective of these imagers is to support the operation of the rover during its drive across the surface (Maki et al., 2012). In particular, the Navigation Cameras are used to monitor the terrain surrounding the vehicle and to perform stereo processing of the retrieved observations, in order to derive surface range maps for hazard detection and target designation purposes. The MSL Navcams are located at the remote sensing mast and are build-to-print copies of the MER mission cameras (Maki et al., 2003). They have a 45-by-45 degree field of view and are equipped with a 1024x1024 pixel CCD detector and a broadband visible filter with an effective wavelength of 650 nm. All the relevant information regarding the performance of the electronics and optics of the engineering cameras can be found in Maki et al., (2003, 2012).

2.1 Image sequences
The MSL Navcam image database has accumulated more than 70,000 images up to mission sol 1648, covering Martian Years 31 to 33. Within the dataset, 7,000 pictures were obtained with the camera pointing upwards, with an instrument elevation angle greater than 10 degrees, so part of the sky was captured. From these Navcam sky observations, we have only considered those on which the solar disc was totally contained within the field of view of the cameras and had a Sun elevation angle greater than 30º. This constraint was set in order to reduce the sensitivity to the vertical distribution of dust in our plane-parallel atmosphere model (Lemmon et al., 2015). This resulted in a final set of 65 images (Figure 1), which formed part of the Surface Attitude Pointing and Positioning (SAPP) sequence used for the calculation of the rover orientation (Maki et al., 2003).

The Navcam observations used in this study are listed in Table 1. For each of these observations, we retrieved from NASA's Planetary Data System (PDS) imaging node the relevant Engineering Data Record (EDR) file, corresponding to the non-processed binary data record produced by the instrument. For complete specifications on the data records, see the MSL Software Interface Specification document (Alexander and Deen, 2017).

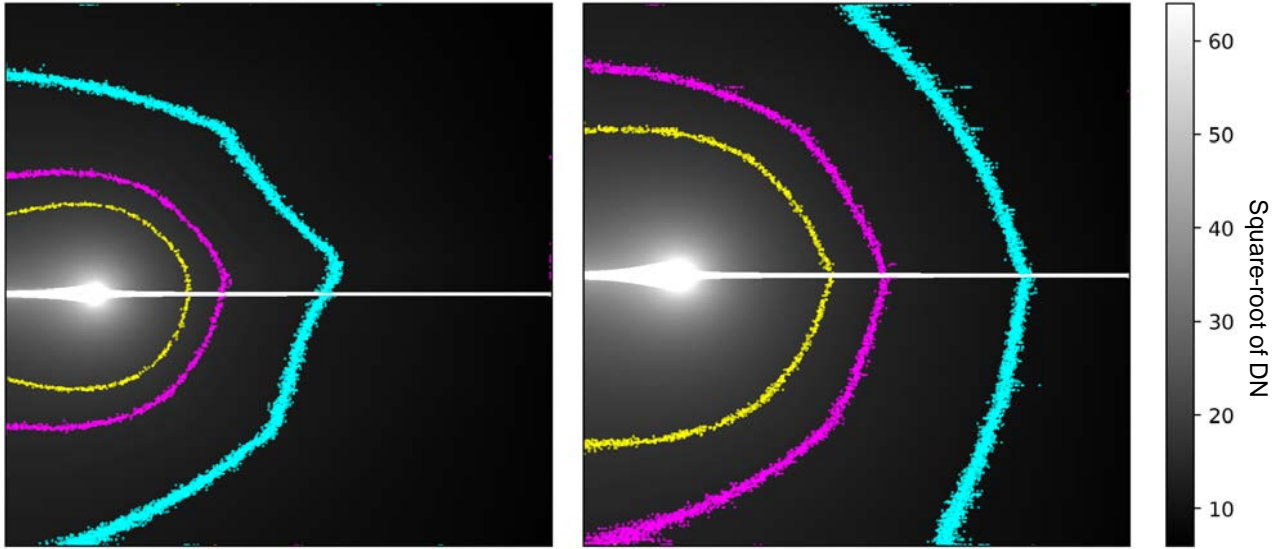

*Figure 1.* MSL Navcam Sun-pointing observations. Images generated from the raw EDR files of the PDS imaging node, with 12-bit resolution (0-4095 DN) and full-frame 45-by-45 degrees FOV. Images are shown in square-root scale. Left: Sol 637 ($L_s$ = 134.4º) with local true solar time (LTST) 13:41:48. Right: Sol 864 ($L_s$ = 269.7º), LTST 13:48:01. On both observations the solar elevation angle above the local horizon was around 56º. The regions with pixel brightness DN values of 100 (cyan), 200 (magenta) and 300 (yellow) have been contoured on-top of these images, showing a clear seasonal variation of the sky brightness as a function of the distance to the Sun. The effects of image smear and blooming on these non-calibrated images can be also appreciated as the solar disc was captured within the frame, as described in Peters (2016).

Table 1. MSL Navcam images used in this study

| MSL SOL | Solar Longitude [deg] | Martian Year | FILENAME | Local True Solar Time | Sun Azimuth [deg] | Sun Elevation [deg] |
|---|---|---|---|---|---|---|
| 21 | 162.0 | 31 | NLA_399363597EDR_F0030100SAPP07612M1.IMG | 14:53:49 | 285.54 | 44.95 |
| 24 | 163.6 | 31 | NLA_399626441EDR_F0030372SAPP07712M1.IMG | 13:57:47 | 291.26 | 58.42 |
| 29 | 166.4 | 31 | NLA_400069880EDR_F0030888SAPP07712M1.IMG | 13:51:24 | 290.26 | 60.30 |
| 39 | 172.0 | 31 | NLA_400958457EDR_F0040468SAPP07712M1.IMG | 14:06:02 | 283.81 | 57.50 |
| 41 | 173.1 | 31 | NLA_401136867EDR_F0041238SAPP07712M1.IMG | 14:20:13 | 281.55 | 54.16 |
| 43 | 174.2 | 31 | NLA_401316500EDR_F0042002SAPP07612M1.IMG | 14:54:12 | 278.36 | 45.91 |
| 48 | 177.1 | 31 | NLA_401761194EDR_F0042644SAPP07612M1.IMG | 15:07:57 | 275.98 | 42.69 |
| 52 | 179.4 | 31 | NLA_402115185EDR_F0043200SAPP07612M1.IMG | 14:50:16 | 275.41 | 47.20 |
| 57 | 182.3 | 31 | NLA_402562341EDR_F0043520SAPP07612M1.IMG | 15:43:47 | 271.94 | 34.01 |
| 102 | 209.5 | 31 | NLA_406558419EDR_F0050388SAPP07612M1.IMG | 16:00:30 | 258.66 | 30.16 |
| 122 | 222.2 | 31 | NLA_408330384EDR_F0050938SAPP07612M1.IMG | 14:57:39 | 251.12 | 44.86 |
| 147 | 238.2 | 31 | NLA_410544820EDR_F0051902SAPP07612M1.IMG | 13:26:20 | 229.73 | 63.29 |
| 166 | 250.6 | 31 | NLA_412233457EDR_F0052330SAPP07612M1.IMG | 13:45:56 | 230.73 | 58.16 |
| 324 | 346.5 | 31 | NRB_426264304EDR_F0060864SAPP07612M1.IMG | 14:03:12 | 266.65 | 59.31 |
| 333 | 351.3 | 31 | NRB_427068209EDR_F0070438SAPP07612M1.IMG | 15:25:13 | 269.00 | 38.83 |
| 340 | 355.0 | 31 | NRB_427685406EDR_F0081148SAPP07612M1.IMG | 14:18:26 | 272.96 | 55.36 |
| 344 | 357.0 | 31 | NRB_428038027EDR_F0090770SAPP07612M1.IMG | 13:39:20 | 276.98 | 64.97 |
| 349 | 359.7 | 31 | NRB_428490085EDR_F0100746SAPP07612M1.IMG | 15:53:35 | 272.68 | 31.50 |
| 358 | 4.2 | 32 | NRB_429282728EDR_F0110882SAPP07612M1.IMG | 14:13:42 | 280.16 | 55.99 |
| 369 | 9.7 | 32 | NRB_430259833EDR_F0120982SAPP07612M1.IMG | 14:26:51 | 282.89 | 52.29 |
| 372 | 11.2 | 32 | NRB_430521464EDR_F0131212SAPP07612M1.IMG | 13:11:46 | 297.40 | 69.77 |
| 383 | 16.6 | 32 | NRB_431506535EDR_F0141428SAPP07612M1.IMG | 15:34:26 | 281.96 | 35.24 |
| 390 | 20.0 | 32 | NRB_432125040EDR_F0151762SAPP07612M1.IMG | 14:49:40 | 287.04 | 45.74 |
| 406 | 27.7 | 32 | NRB_433539267EDR_F0162120SAPP07612M1.IMG | 13:15:33 | 310.60 | 65.29 |
| 412 | 30.5 | 32 | NRB_434076977EDR_F0171310SAPP07612M1.IMG | 14:39:55 | 293.75 | 46.74 |
| 419 | 33.8 | 32 | NRB_434700865EDR_F0181406SAPP07612M1.IMG | 15:22:30 | 290.85 | 36.52 |
| 426 | 37.1 | 32 | NRB_435323449EDR_F0191256SAPP07612M1.IMG | 15:43:53 | 290.91 | 31.22 |
| 433 | 40.3 | 32 | NRB_435936028EDR_F0201326SAPP07612M1.IMG | 13:22:53 | 315.58 | 60.92 |
| 440 | 43.5 | 32 | NRB_436559934EDR_F0211648SAPP07612M1.IMG | 14:05:36 | 305.72 | 52.15 |
| 454 | 49.9 | 32 | NRB_437800741EDR_F0221028SAPP07612M1.IMG | 13:37:08 | 315.40 | 56.37 |

| MSL SOL | Solar Longitude [deg] | Martian Year | FILENAME | Local True Solar Time | Sun Azimuth [deg] | Sun Elevation [deg] |
|---|---|---|---|---|---|---|
| 470 | 57.2 | 32 | NRB_439224987EDR_F0231524SAPP07612M1.IMG | 14:44:26 | 303.95 | 42.25 |
| 494 | 67.9 | 32 | NRB_441349865EDR_F0240562SAPP07612M1.IMG | 13:18:00 | 326.49 | 56.24 |
| 527 | 82.8 | 32 | NRB_444282928EDR_F0251906SAPP07612M1.IMG | 14:21:28 | 312.50 | 44.63 |
| 545 | 90.9 | 32 | NRB_445883715EDR_F0261458SAPP07612M1.IMG | 15:10:31 | 305.31 | 34.98 |
| 552 | 94.1 | 32 | NRB_446499987EDR_F0271500SAPP07612M1.IMG | 13:48:00 | 319.92 | 50.32 |
| 563 | 99.1 | 32 | NRB_447479279EDR_F0281504SAPP07612M1.IMG | 14:34:31 | 310.06 | 42.26 |
| 569 | 101.9 | 32 | NRB_448010997EDR_F0291606SAPP07612M1.IMG | 14:20:15 | 312.31 | 45.07 |
| 589 | 111.2 | 32 | NRB_449782783EDR_F0301366SAPP07612M1.IMG | 13:22:50 | 325.09 | 55.44 |
| 631 | 131.4 | 32 | NRB_453512631EDR_F0311670SAPP07612M1.IMG | 13:50:16 | 311.36 | 54.24 |
| 637 | 134.4 | 32 | NRB_454044666EDR_F0321252SAPP07612M1.IMG | 13:41:18 | 312.45 | 56.48 |
| 647 | 139.5 | 32 | NRB_454932444EDR_F0331334SAPP07612M1.IMG | 13:43:25 | 309.53 | 57.09 |
| 657 | 144.6 | 32 | NRB_455821590EDR_F0341616SAPP07612M1.IMG | 14:07:46 | 301.35 | 53.10 |
| 662 | 147.2 | 32 | NRB_456267998EDR_F0351626SAPP07612M1.IMG | 14:49:43 | 293.68 | 44.20 |
| 668 | 150.3 | 32 | NRB_456799315EDR_F0361708SAPP07612M1.IMG | 14:29:06 | 294.75 | 49.31 |
| 672 | 152.4 | 32 | NRB_457152376EDR_F0371824SAPP07612M1.IMG | 13:56:42 | 299.06 | 56.85 |
| 685 | 159.4 | 32 | NRB_458310507EDR_F0381758SAPP07612M1.IMG | 15:04:35 | 285.99 | 42.12 |
| 705 | 170.5 | 32 | NRB_460084993EDR_F0391930SAPP07612M1.IMG | 14:50:54 | 280.83 | 46.44 |
| 733 | 186.7 | 32 | NRB_462572793EDR_F0402484SAPP07612M1.IMG | 15:27:01 | 270.07 | 38.33 |
| 747 | 195.0 | 32 | NRB_463813017EDR_F0412270SAPP07612M1.IMG | 14:44:14 | 265.70 | 49.10 |
| 864 | 269.8 | 32 | NRB_474199389EDR_F0443000SAPP07612M1.IMG | 13:48:01 | 228.85 | 56.93 |
| 952 | 323.9 | 32 | NRB_482013079EDR_F0452302SAPP07612M1.IMG | 13:35:01 | 245.47 | 64.61 |
| 964 | 330.8 | 32 | NRB_483083182EDR_F0462052SAPP07612M1.IMG | 14:52:45 | 257.44 | 46.69 |
| 984 | 341.9 | 32 | NRB_484852932EDR_F0471818SAPP07612M1.IMG | 13:21:25 | 260.64 | 69.55 |
| 1067 | 24.1 | 33 | NRB_492221950EDR_F0482954SAPP07612M1.IMG | 13:58:10 | 296.53 | 57.13 |
| 1104 | 41.4 | 33 | NRB_495505377EDR_F0493088SAPP07612M1.IMG | 13:51:37 | 307.87 | 55.32 |
| 1167 | 69.9 | 33 | NRB_501099548EDR_F0503368SAPP07612M1.IMG | 14:32:05 | 308.98 | 43.42 |
| 1262 | 113.3 | 33 | NRB_509536696EDR_F0523240SAPP07612M1.IMG | 15:45:07 | 299.38 | 28.53 |
| 1301 | 132.2 | 33 | NRB_512994632EDR_F0533062SAPP07612M1.IMG | 14:41:32 | 301.25 | 44.02 |
| 1376 | 171.9 | 33 | NRB_519653718EDR_F0543156SAPP07612M1.IMG | 15:08:58 | 278.92 | 42.11 |
| 1433 | 205.6 | 33 | NRB_524708540EDR_F0562614SAPP07612M1.IMG | 13:42:01 | 255.24 | 64.04 |
| 1468 | 227.8 | 33 | NRB_527820348EDR_F0573480SAPP07612M1.IMG | 14:47:57 | 247.96 | 46.77 |
| 1503 | 250.5 | 33 | NRB_530924083EDR_F0583228SAPP07612M1.IMG | 13:33:52 | 227.71 | 60.45 |
| 1571 | 294.1 | 33 | NRB_536961350EDR_F0593184SAPP07612M1.IMG | 12:59:09 | 216.75 | 66.85 |
| 1604 | 314.2 | 33 | NRB_539892888EDR_F0603516SAPP07612M1.IMG | 13:19:28 | 234.43 | 66.55 |
| 1646 | 338.3 | 33 | NRB_543622726EDR_F0613478SAPP07612M1.IMG | 13:39:02 | 258.44 | 65.03 |

*Note: MSL mission sol number, solar longitude (in degrees), Martian Year (following Clancy et al. 2000 convention), local true solar time, Sun azimuth and elevation angles (in degrees) are given with respect to North and local horizon. These values were used as input parameters for the models.*

2.2 Photometric calibration

The MSL Navcam raw EDR files with 12-bit pixel DN values were converted into physical units of absolute radiance (W m$^{-2}$ nm$^{-1}$ sr$^{-1}$). As the MSL Engineering Cameras (Navcam, Hazcam) are built-to-print copies of the MER Engineering Cameras (Maki et al., 2012), for this study we followed the calibration process described in Section 2 of Soderblom et al., (2008a) for the MER Navcam in-flight data.

On the lines below we summarise the calibration steps for the MSL Navcam EDR files. The values of the calibration parameters customized for MSL rover are provided on Table 2. We refer to Soderblom et al., (2008a) for a comprehensive and detailed description of the rover engineering cameras in-flight data calibration procedure.

For an MSL Navcam observation EDR file, the following corrections were applied through the calibration process to a pixel located at row *i* and column *j*:

$$C(i,j) = \frac{\left(R(i,j) - B(j, T_{Elec}) - D(i,j, T_{CCD}, t_{exp}) - S(j)\right)}{F(i,j)} \cdot \frac{1}{t_{exp}} \qquad (1)$$

Here C is the flux value for the calibrated image pixel in units of DN/s, R is the raw EDR input value of the pixel in DN, B is the bias correction, depending on the electronics temperature $T_{Elec}$ (°C) and the column of the image, D is the dark current correction depending on the exposure time $t_{exp}$ (seconds) and the temperature of the imager's CCD $T_{CCD}$ (ºC), S is the shutter smear removal and F is the flat field correction.

*1. Bias removal.* On the edges of the Navcam detectors there are 16 pixels ("reference pixels", files coded "ERP" in the PDS archive) that record the bias added by the video offset to the signal to prevent it from reaching zero values. Within an ideal scenario, these reference pixel files shall be obtained for each observation so the added bias could be estimated and subtracted. However, due to downlink data-rate limitations, there are only few images with reference pixel data available (for MSL Navcam, up to sol 1648 and out of ~70,000 observations, there were only 520 reference pixel files available at the PDS archive). The bias was modelled into two components, a mean bias and an image-line position dependant offset:

$$B(j, T_{Elec}) = a_0 + a_1 \exp(a_2 T_{Elec}) + \Delta bias + bias\_offset(j) \qquad (2)$$

In this work, we modelled the MSL Navcam bias using the available reference pixel files for each camera; the offset voltage *Δbias = (4095 – video_offset ) / 2* is equal to zero as the video_offset parameter was set to 4095 for all the observations evaluated in this study. For each Navigation camera, an individual set of parameters (*a0, a1*) were derived; while the image-line number dependant offset were approximated as a logarithmic function in the form of *bias_offset(j) = -1.85 + 0.31 log(j)*, in units of DN and for *j* from 1 to 1024.

2. *Dark current removal.* The frame transfer readout method of the Navigation Cameras allows the modelling of the dark current into two separate components, the active area and the masked area dark current contributions. They correspond to the contributions of the accumulated charge when the detector is exposed to the scene or when it is being read out, respectively. In both cases, the dark current was modelled with a mean rate in the form of an exponential function of the CCD temperature multiplied by a scaling factor contained in the masked and active area dark flats.

$$D(i,j,T_{CCD},t_{exp}) = D_{Masked}(i,j,T_{CCD}) + D_{Active}(i,j,T_{CCD},t_{exp}) \qquad (3)$$
$$D_{Masked}(i,j,T_{CCD}) = c_0 \cdot \exp(c_1 T_{CCD}) \cdot DF_{Masked}(i,j) \qquad (3.1)$$
$$D_{Active}(i,j,T_{CCD},t_{exp}) = t_{exp} \cdot d_0 \cdot \exp(d_1 T_{CCD}) \cdot DF_{Active}(i,j) \qquad (3.2)$$

The masked and active dark flats (*$DF_{Masked}$, $DF_{Active}$*) were built from the averaging and normalisation of zero and non-zero exposure time dark images, respectively.

For this study, we generated these masked and active dark flats using the available dark images, which were obtained during the cruise stage of the MSL mission and archived in the correspondent folder within the PDS. The dark current rate modelling parameters were then retrieved following the indications provided in Section 2.4 of Soderblom et al., (2008a).

*3. Shutter smear removal.* With no mechanical shutter, there is an accumulation of additional charge at the detector's active region during the data transfer (Figure 1). This additional scene-dependent charge needs to be calculated and removed from the image in a recursive manner for each image line starting from the closest line to the read-out region.

The equations in Section 5.2 of Soderblom et al., (2008a) were used for the modelling and removal of the shutter smear effect in the MSL Navcam images, taking into consideration the particular frame transfer direction of each camera (Peters, 2016)

*4. Flat field correction.* This step corrects the variations in pixel-to-pixel responsivity using previously retrieved uniformly illuminated images. The pre-flight flat images are located in the MSL mission archive of the PDS Imaging Node for each navigation camera.

*5. Conversion to physical units.* The transformation from the resulting Navcam calibrated image flux $C_{ij}$ (in units of DN/s) to a calibrated image of the scene with physical units of absolute radiance $L_{ij}$ (units: W m$^{-2}$ nm$^{-1}$ sr$^{-1}$) was modelled with a linear equation dependent on the camera's CCD temperature in the form of:

$$L(i,j) = (K_0 + K_1 T_{CCD}) \cdot C(i,j) \qquad (4)$$

With offset and slope coefficients $K_0$ (units: DN/s) and $K_1$ (units: (DN s$^{-1}$) ºC$^{-1}$).

For MSL Navigation Cameras, the radiometric conversion coefficients for each imager were not available; we estimated the values for $K_0$ and $K_1$ from the default radiometric conversion coefficients of MER Navcams by averaging the coefficients provided for those cameras in Table 2 of Soderblom et al., (2008a).

Table 2. MSL Navigation Camera calibration parameters

| CALIBRATION STAGE | MSL NAVIGATION CAMERAS | | | | SOURCE |
|---|---|---|---|---|---|
| | **NAV_RIGHT_A**: SN_0206 | **NAV_RIGHT_B**: SN_0218 | **NAV_LEFT_A**: SN_0216 | **NAV_LEFT_B**: SN_0215 | |
| Bias removal | a0 = -176.64 DN, a1 = 190.5 DN, a2 = 0.0033 ºC$^{-1}$ | a0 = -30.46 DN, a1 = 41.3 DN, a2 = 0.0125 ºC$^{-1}$, | a0 = -37.59 DN, a1 = 51.5 DN, a2 = 0.0095 ºC$^{-1}$, | a0 = -10.43 DN, a1 = 45.4 DN, a2 = 0.0108 ºC$^{-1}$, | Derived for MSL |
| Dark current removal: parameters | Masked region mean rate: c0 = 4.155 DN; c1 = 0.1112 ºC$^{-1}$ Active region mean rate: d0 = 12.096 DN, d1 = 0.1010 ºC$^{-1}$ | | | | Derived for MSL, PDS [(1)] |
| Masked dark flat image | NRA_384856702EDR_F, NRA_384856709EDR_F | NRB_388221633EDR_F, NRB_388221626EDR_F | NLA_384856702EDR_F, NLA_384856709EDR_F | NLB_388221626EDR_F, NLB_388221633EDR_F | PDS [(1)(2)] |
| Active dark flat image | NRA_384856744EDR_F, NRA_384856717EDR_F | NRB_388221640EDR_F, NRB_388221668EDR_F | NLA_384856717EDR_F, NLA_384856744EDR_F | NLB_388221640EDR_F, NLB_388221668EDR_F | PDS [(1)(2)] |
| Flat field correction | MSL_FLAT_SN_0206.IMG | MSL_FLAT_SN_0218.IMG | MSL_FLAT_SN_0216.IMG | MSL_FLAT_SN_0215.IMG | PDS [(3)] |
| Conversion to physical units | K0 = 9.634e-6 W m$^2$ nm$^{-1}$ sr$^{-1}$ (DN s$^{-1}$)$^{-1}$ ; K1 = 1.035e-8 W m$^2$ nm$^{-1}$ sr$^{-1}$ (DN s$^{-1}$)$^{-1}$ ºC$^{-1}$ | | | | Adapted from MER [(4)] |

(1): https://pds-imaging.jpl.nasa.gov/data/msl/MSLNAV_0XXX/DATA/CRUISE/
(2): Dark masked and active flats are available in this public repository: http://www.ajax.ehu.es/hcc/Icarus2018153/
(3): https://pds-imaging.jpl.nasa.gov/data/msl/MSLNAV_0XXX/CALIB/
(4): Table 2 from Soderblom et al., (2008a). A 15% of uncertainty is assumed for K0 and K1 values.

From the calibration parameters listed above, a noticeable deviation in the bias parameters for the Navcam Right A (SN_0206) can be identified when compared to the other cameras. It is worth mentioning that the last observation of this imager available in the PDS corresponds to MSL mission sol 199 and no data from this camera were used in this work.

The uncertainty of the calibration procedure was estimated by comparing the radiance of Navcam calibrated images with the radiometrically calibrated data of MSL Mastcam (Bell et al., 2017). The retrievals of Mastcam-Left Filter number 4 were used, as the effective wavelength of this filter (674 nm) is the closest one to the effective wavelength of the navigation cameras (~ 650 nm) (Maki et al., 2012). Mastcam data were obtained from the PDS and the conversion from the archived 12-bit DN pixel values to units of radiance factor (I/F) and absolute radiance (W m$^{-2}$ nm$^{-1}$ sr$^{-1}$) was done as described in Section 5.2.7 of Bell et al., (2017).

Martian sky and surface observations with both cameras capturing the same scene and similar pointing, retrieved during the same sol at an approximate local true solar time were selected, ending up with a total of 16 pairs of Navcam and Mastcam images (Table 3). Several regions of interest appearing on both observations were then chosen and the mean radiance value of the regions was obtained and compared (Figure 2). This was evaluated for around 110 different regions within the 16 Navcam-Mastcam pairs of image data

This comparison showed that the radiance value differences between the calibrated MSL Navcam images and the Mastcam radiometrically corrected data were less than 2%. This result is of the same order as the obtained by Soderblom et al., (2008a) for MER Navcams when compared, in that case, to the MER Panoramic Camera (Pancam). As the absolute radiance uncertainty for MSL Mastcam was estimated of the order of 10% in Bell et al. (2017), we considered that the absolute radiance uncertainty for the Navcam images calibrated in this study is about 12%. In addition to this uncertainty, the pixel-to-pixel precision was also evaluated by calculating the variance of the radiance values of the gray-scale rings of the Mastcam calibration target, which is considered as reflectance-uniform for observations taken near to the rover landing date (Soderblom et al., 2008a). We used the Navcam image of the calibration target obtained on sol 71 and the estimated relative pixel-to-pixel precision values were around 3.5%.

Finally, due to the nature of the observations (Sun-pointing images), the effects of internal scattered instrumental light had to be also considered. Pre-flight (Maki et al., 2004) and in-flight (Soderblom et al., 2007) stray light tests were performed for MER navigation cameras, showing almost no internal instrument debilitating signal, and providing reliable sky radiances down to distances of around 2º from the Sun (Soderblom et al., 2008a; Smith and Wolff, 2014). For the MSL Navcam SAPP images used in this work, a qualitative characterisation of the stray and scattered light effect was performed. The sky radiances of the calibrated images were evaluated as a function of the distance to the solar disc centre. The analysis of these sky brightness radial profiles showed no evidence of clear ghosts or glints on the retrieved data in the non-saturated pixel near Sun region at distances > 4º (see Fig. 3).

Table 3. MSL Navcam and Mastcam observation pairs for validating the calibration

| Navcam file | Sol | LTST | Mastcam File | Sol | LTST | Pointing |
|---|---|---|---|---|---|---|
| NLA_400157480EDR_D0040000NCAM00510M1.IMG | 30 | 13:32:28 | 0030ML0001340020100850D01_DRXX.IMG | 30 | 13:36:12 | Sky |
| NLA_400791104EDR_F0040000NCAM00514M1.IMG | 37 | 16:51:14 | 0037ML0001640030101309D01_DRXX.IMG | 37 | 16:48:31 | Sky |
| NLA_403614285EDR_D0050104NCAM00524M1.IMG | 69 | 12:07:28 | 0069ML0004860040102539D01_DRXX.IMG | 69 | 12:17:19 | Ground |
| NLA_403797280EDR_F0050104NCAM00526M1.IMG | 71 | 13:35:47 | 0071ML0004980040102589D01_DRXX.IMG | 71 | 13:32:06 | Cal.targ |
| NLB_421372569EDR_F0060000NCAM00101M1.IMG | 269 | 11:33:52 | 0269ML0011790040106119D01_DRXX.IMG | 269 | 11:36:25 | Ground |
| NLB_449260422EDR_M0300786NCAM00505M1.IMG | 583 | 16:08:54 | 0583ML0024390370300420D01_DRXX.IMG | 583 | 16:01:10 | Sky |
| NLB_452004100EDR_F0311330NCAM00322M1.IMG | 614 | 13:58:01 | 0614ML0025940050301802D01_DRXX.IMG | 614 | 14:02:43 | Ground |
| NRB_452518799EDR_F0311330NCAM00323M1.IMG | 620 | 09:07:45 | 0620ML0026540020302355D01_DRXX.IMG | 620 | 09:03:50 | Cal.targ |
| NLB_461944914EDR_F0401378NCAM00390M1.IMG | 726 | 13:43:05 | 0726ML0031010050305083D01_DRXX.IMG | 726 | 13:40:12 | Ground |
| NLB_462486418EDR_D0402040NCAM00556M1.IMG | 732 | 16:05:57 | 0732ML0031410080205207D01_DRXX.IMG | 732 | 16:00:44 | Sky |
| NLB_468598450EDR_F0441140NCAM02343M1.IMG | 801 | 12:15:14 | 0801ML0034990020400821D01_DRXX.IMG | 801 | 12:24:47 | Cal.targ |
| NLB_505708078EDR_F0520936NCAM00203M1.IMG | 1219 | 12:35:44 | 1219ML0055920120503562D01_DRXX.IMG | 1219 | 12:33:47 | Ground |
| NLB_508102653EDR_F0521370NCAM00320M1.IMG | 1246 | 12:01:32 | 1246ML0058130120504007D01_DRXX.IMG | 1246 | 12:08:05 | Ground |
| NLB_509965530EDR_F0530186NCAM00320M1.IMG | 1267 | 11:41:51 | 1267ML0059320120504318D01_DRXX.IMG | 1267 | 11:54:30 | Ground |
| NLB_511122556EDR_F0531182NCAM00320M1.IMG | 1280 | 12:31:39 | 1280ML0060170120504773D01_DRXX.IMG | 1280 | 12:36:42 | Ground |
| NLB_521958717EDR_M0052444NCAM00567M1.IMG | 1402 | 14:19:41 | 1402ML0068710030601789D01_DRXX.IMG | 1402 | 14:10:13 | Sky |

*Notes: The camera pointing is provided in the last column. "Cal.targ" stands for the Mastcam Calibration Target, mounted at the right-side of the top rover's deck (Bell et al., 2017).*

2.3 Geometric reduction

The geometric reduction of the near-Sun images was performed using the CAHVOR photogrammetric camera model system (Yakimovsky and Cunningham, 1978; Gennery, 2006), as described in Maki et al. (2012).

In this camera system a 3-dimensional point in the scenery is transformed into image pixel row-column coordinates using a system of six vectors: the camera centre position (C) and unit perpendicular axis (A) vectors, the horizontal (H) and vertical (V) information vectors, and the optical (O) and radial distortion (R). The component values of these vectors were retrieved from each of the observations' PDS label and can be inverted in order to assign to each image pixel the corresponding values of azimuth and elevation (Di and Li, 2004; Gennery, 2006) in the site coordinate frame system, with positive X, Y, and Z axes pointing at Mars' North, East and gravity nadir, respectively (Alexander and Deen, 2017). These values were then used together with the solar site azimuth and elevation angles to derive the relevant scattering angle for each pixel (Figure 3).

When performing the geometric reduction of the Navcam Sun viewing image data, it was noticed that the label recorded Sun centre position coordinates presented some drift (generally less than 1º degree) from the actual solar disc centre on the image due to the rover's attitude at the time of the observation, as it has been also stated for MER (e.g., Soderblom et al., 2008a, Lemmon et al., 2015). For these cases, the centre of the bright disc was located, the azimuth and elevation angles were derived and compared against the labelled Sun position; when there was a drift of more than 0.25º, the Sun position was updated and the scattering angle for each pixel was then re-calculated.

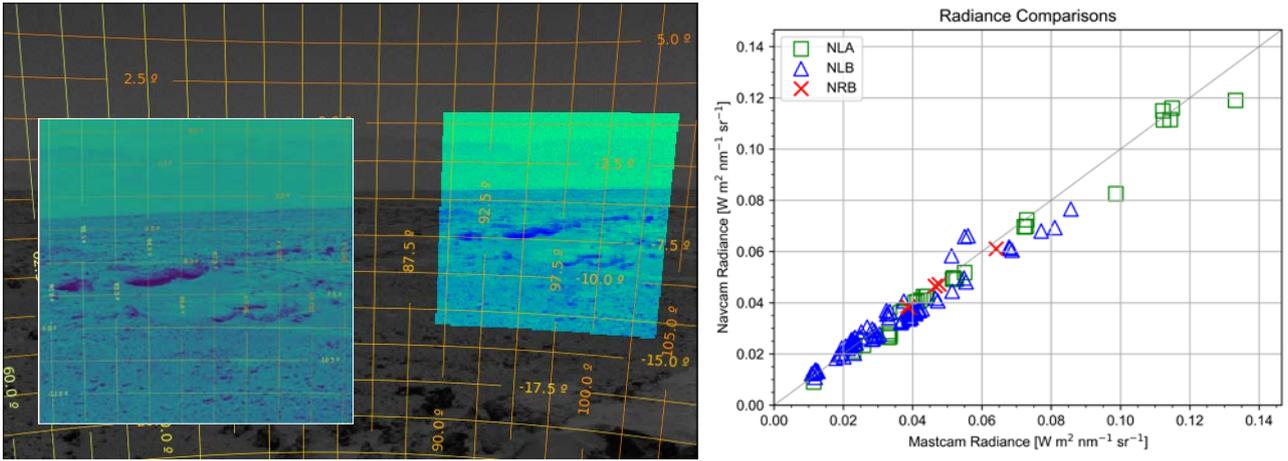

*Figure 2. MSL Navcam calibration validation. (Left) An example of a matching-pair of Navcam-Mastcam images used for the calibration comparison is provided. The Navcam observation (NLA_403614285EDR_D0050104NCAM00524M1.IMG) was obtained on sol 69 around local noon (12:07:28 LTST) with a solar longitude of 189.3 degrees. Calibration and geometric reduction was performed on the file, and the grid indicating elevation and azimuth in the rover coordinate frame is provided. The area highlighted in blue at the right of the image shows the region captured by the matching Mastcam image for this observation, which is provided on the left-side inset, and corresponds to the file 0069ML0004860040102539D01_DRXX.IMG, obtained ten minutes after the Navcam observation (12:17:19 LTST), on the same sol. (Right) The absolute radiance for several regions of interest (approx. 110) within the 16 pair of Navcam-Mastcam matching-observations were retrieved and compared, resulting in mean radiance difference of less than 2%.*

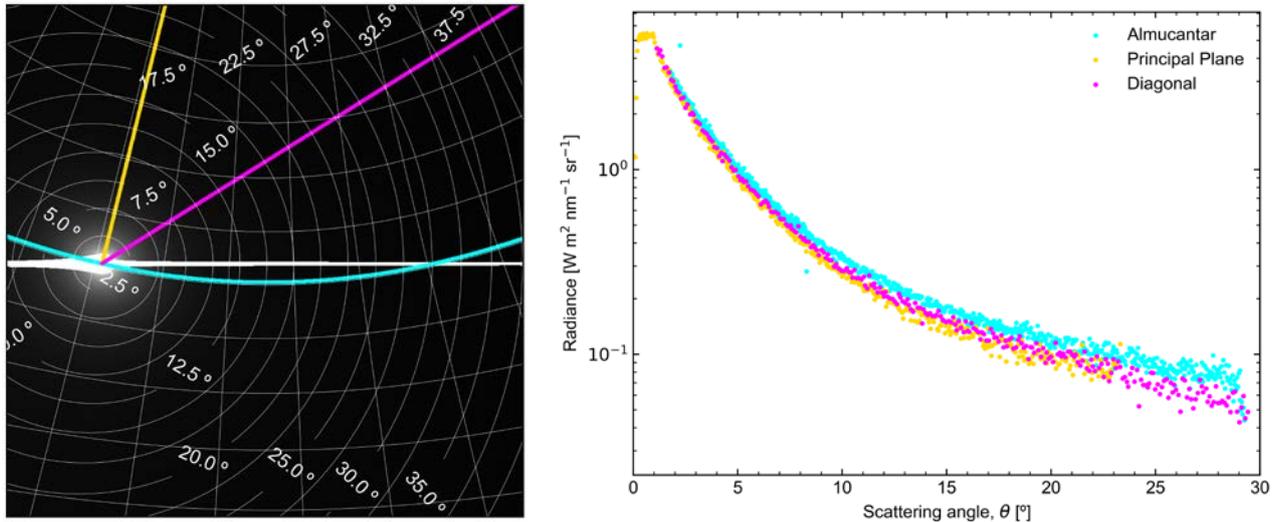

*Figure 3. Retrieval of sky brightness curves from Navcam's observation. (Left) Navcam image file NRB_519653718EDR_F0543156SAPP07612M1.IMG, obtained at LTST 15:08:58 on sol 1376 ($L_s$ = 171.85º). Sun elevation and azimuth angles were 42.11º and 278.92º, respectively, in local site frame. Calibration and geometric reduction were performed on the image data. Saturated pixels (white region) were masked off the image and the azimuth-elevation grid and scattering angle contour lines are provided. The sampling paths for different directions are indicated: almucantar (in cyan, along the Sun's elevation angle), principal plane (in yellow, along the Sun's azimuth) and diagonal (in magenta). (Right) The sky brightness as a function of the scattering angle for each sampling direction.*

## 3. Modelling and methodology

The airborne aerosol properties of Mars' atmosphere were characterised by comparing the MSL Navcam retrievals of the near-Sun sky brightness with model computations. In the following paragraphs we describe the procedure we followed to simulate the radiance factor (*I/F*) observed by the Navigation Cameras and the method used to retrieve the best fitting parameters.

3.1 Radiative transfer
We used the discrete ordinates method (Stamnes et al., 1988) to solve the radiative transfer equation with multiple scattering in a plane-parallel atmosphere in order to model the sky radiances as a function of the scattering angle. We used a Python implementation (PyDISORT, Ádámkovics et al., 2016) of the version 2.1 of DISORT, translated from FORTRAN into C (CDISORT, Buras et al., 2011).

3.2 Atmosphere structure and composition
The atmosphere above Gale Crater was modelled with 30 plane-parallel layers distributed vertically in linearly spaced pressure levels with a total height of 100 km. For each layer, the values for the atmospheric pressure, temperature, density and composition were retrieved from the Mars Climate Database (MCD, v.5.2) (Forget et al., 1999; Millour et al., 2015). These variables were loaded from the database as a function of the observation's local true solar time (LTST), solar longitude ($L_s$) and location (Mars' Gale Crater: 4.6ºS; 137.4ºE) and interpolated at each layer.

Martian atmospheric constituent species considered in our model ($CO_2$, $H_2O$, $O_2$, $N_2$, and $O_3$) presented no strong gas absorption within the Navcam wavelength band (600 to 800 nm), so their contribution to the atmospheric opacity was considered negligible. For the Rayleigh scattering by gases, only the contribution of the $CO_2$ was taken into account in this study. The Rayleigh scattering cross section was obtained using the model and constants in Sneep and Ubachs (2005).

The aerosol optical depth at each atmospheric layer was set using a modified Conrath profile (Forget et al., 1999). This profile models the vertical distribution of the aerosol mass mixing ratio, when integrated from the top of the atmosphere to a specific height level (Heavens et al., 2011), an expression for the column optical depth *τ* is obtained in the form of:

$$\tau(z) = \tau_0 \cdot \tilde{\sigma}(z) \cdot \exp\left[\nu \cdot (1 - \tilde{\sigma}(z)^{-l})\right] \qquad (5)$$

where $\tau_0$ is a reference optical depth at the surface, $\tilde{\sigma}$ is the ratio between the pressure *p* at a specific level and a reference pressure level $p_0$, *l* is the ratio between a reference height and the maximum altitude of observed dust $z_{max}$, and the parameter *ν* is the ratio between the dust diffusion and surface sedimentation characteristic times (Conrath, 1975). By means of observational data, it was derived that for a reference height of *70 km*, the value for this parameter is *ν = 0.007* and the $z_{max}$ depends on the latitude and solar longitude (Forget et al., 1999).

3.3 Aerosol model
The radiative transfer code required only 3 parameters at each layer of the discretised atmosphere model for defining the airborne aerosol: the single scattering albedo ($\omega_0$), the normalised phase function *P(θ)*, where *θ* is the scattering angle, and the optical depth *τ* at the specific layer. The aerosol's single scattering albedo and phase function were retrieved with a T-Matrix method (e.g., Mishchenko and Travis, 1998) using a FORTRAN code (https://www.giss.nasa.gov/staff/mmishchenko/t_matrix.html) for randomly oriented non-spherical particles. The dust particle refractive indices were interpolated from Wolff et al. (2009) at the Navcam effective wavelength.

We assumed a well mixed dust situation, so the same dust aerosol effective radius and phase function values were considered at each atmospheric layer in the model. Dust aerosol particles were modelled as cylinders with diameter-to-length aspect ratios (D/L) of 1.0 (Wolff et al., 2009), following a log-normal particle size distribution (e.g., Hansen and Travis, 1974). The effective variance $v_{eff}$ was fixed to 0.30 (e.g., Tomasko et al., 1999; Wolff et al., 2009; Smith and Wolff, 2014) and the effective radius $r_{eff}$ (defined as equivalent volume radius) was left as free parameter.

For upward looking observations and at small scattering angles, the single scattering of the particles dominates the atmospheric scattering function (Pollack et al., 1995), so the brightness function retrieved near the Sun has low sensitivity to the surface scattering. A Lambertian surface was assumed in our radiative transfer model with value of 0.20 for the Lambert albedo, corresponding to the average surface albedo at Gale Crater (Anderson and Bell, 2010).

3.4 Retrieval procedure

A brute-force iterative retrieval scheme was implemented based in the comparison of the sky brightness curve obtained from Navcam observations with the modelled curve generated with our radiative transfer code for the 2 free parameters: the particle size distribution effective radius ($r_{eff}$) and the dust column optical depth at surface ($\tau_0$). The retrieval output values for these parameters were those generating the best fitting curve under a lowest mean quadratic deviation $\chi^2$ criterion.

For each Navcam observation we proceeded in the following manner:

1. The EDR raw image file was calibrated according to the method described in Section 2 in order to obtain the observed scene radiance $L_{obs}$ (W m$^2$ nm$^{-1}$ sr$^{-1}$). This radiance was then converted into approximated radiance factor $(I/F)_{obs}$ by dividing each pixel's radiance value by the solar spectral irradiance at the top of the atmosphere at the time of the observation convolved to the Navcam filter bandpass (1.524 W m$^2$ nm$^{-1}$ sr$^{-1}$ at 1 AU), and divided by π (e.g., Soderblom et al., 2008a; Bell et al., 2017). The solar spectral irradiance data was obtained from Colina et al. (1996).

2. Geometric reduction was performed in the calibrated image as per Section 2.3. For each pixel of the image, the corresponding values for the site azimuth and elevation were derived and the scattering angles were calculated.

3. The Navcam observed sky brightness as a function of the scattering angle curve was generated by retrieving the image's sky radiance factor values along a diagonal sampling path (Figure 3). This path started at the centre of the solar disc (scattering angle = 0º) and finished at the furthest sky point, which due to the geometry of the observations was located at the upper right corner of the 1024x1024 pixel image. This sampling direction was selected in order to reduce the importance of the aerosol vertical distribution by avoiding points with low elevation, and cover as much part of the sky brightness curve as possible (Soderblom et al., 2008a; Soderblom et al., 2008b). The retrieved sky radiance curve was sampled from a scattering angle of $\theta$ = 4º to 30º with steps of 1º; this was done in order to skip the saturated pixels in the very near solar disc region and limit the possible contributions from instrumental stray and scattered light into the sampled data. In addition, this also alleviated the computational time requirements related to the number of streams used in the radiative transfer solver scheme.

4. The modelled curve was generated using the radiative transfer model. For the solar longitude ($L_s$) and local true solar time (LTST) of a Navcam observation, the model atmosphere structure was initiated and the atmospheric parameters at each layer were retrieved from the MCD. Dust aerosol radiative properties (single scattering albedo, phase function) were loaded from pre-calculated look-up tables as a function of the aerosol shape (cylinders of aspect ratio D/L = 1), particle size distribution effective radius ($r_{eff}$, free parameter) and effective variance ($v_{eff}$, fixed to 0.3). The vertical profile of the dust optical depth was generated using the expression in [5], which depended on the vertical profile of the atmospheric pressure, the solar longitude and the dust column optical depth at surface, $\tau_0$ (free parameter).

5. Once the model was created, the radiative transfer equation was solved using the discrete ordinates method (DISORT) for each point in the sky along the defined sampling direction, in order to obtain the modelled sky brightness (in radiance factor, $I/F$) as a function of the scattering angle. The viewing geometry configuration in the simulation was defined from the position of the Sun and the sky point coordinates retrieved along the sampling path. The number of moments used in the expansion of the modelled aerosol phase function was set to 250 and the number of streams was fixed to 32.

6. The Navcam observed sky brightness as a function of the scattering angle curve $(I/F)(\theta)_{obs}$ and the modelled curve $(I/F)(\theta)_{model}$ were compared using a standard $\chi^2$ method defined as:

$$\chi^2 = \sum_{i=1}^{N} \left( \frac{(I/F)_{obs_i} - (I/F)_{model_i}}{\sigma_i \cdot (I/F)_{obs_i}} \right)^2 \qquad (6)$$

Where for the *N* sampled points along the curve, the Navcam and model radiance factors at the specific scattering angle were compared using a least squares quadratic error criterion, with the variance $\sigma_i = 0.12$ associated to the absolute calibration uncertainty (12%) of MSL Navcams derived in Section 3. The reduced $\chi^2$ value ($\chi_\nu^2$), corresponded to the obtained $\chi^2$ divided by the number of degrees of freedom $\nu = N - 2$ (number of sampled points minus the number free parameters of the retrieval, $r_{eff}$ and $\tau_0$).

This $\chi^2$ curve-fitting comparison was done in a successive manner for each of the modelled sky brightness curves generated with combinations of the aerosol model free parameter values $r_{eff}$ and $\tau_0$. In order to cover a broad range of possible scenarios, the effective radius was iterated from 0.5 to 2.5 µm with steps of 0.02 µm; and the column aerosol optical depth value at surface was sampled between 0.1 and 2.5 with steps of 0.02. The size of the step was selected due to the computational time and the limits of the sampling region for the $r_{eff} - \tau_0$ space were defined based on the minimum and maximum values retrieved by previous studies at the MSL landing site (e.g., Lemmon, 2014; Smith et al., 2016; Vicente-Retortillo et al., 2017; McConnochie et al., 2017).

7. The set of parameters ($r_{eff}$, $\tau_0$) returning the minimum value for the mapped $\chi^2$ were considered the solutions of the retrieval and the uncertainty level associated to each parameter was calculated from the 68% confidence region (*1σ* error) (Figure 4).

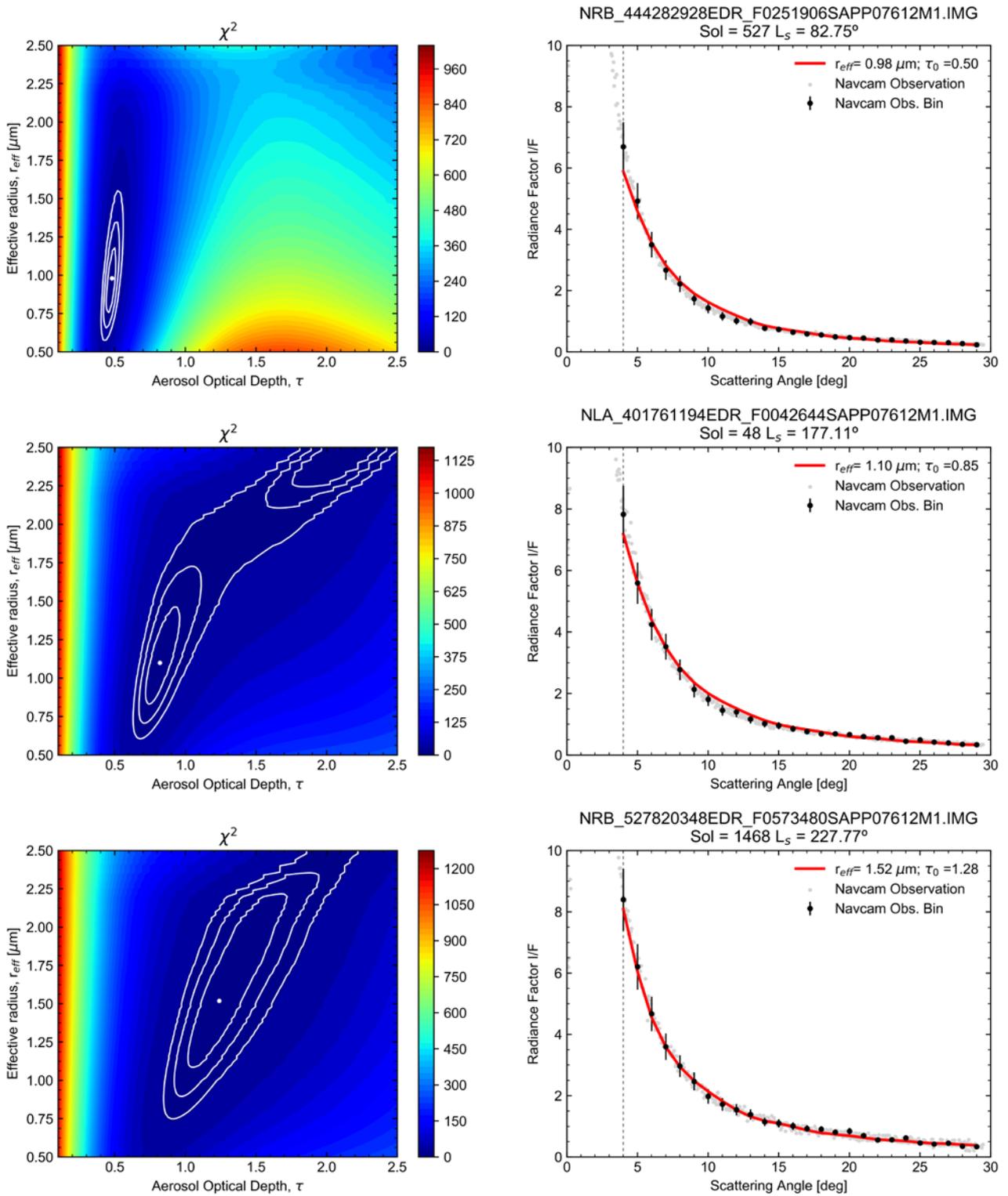

*Figure 4. Effective radius and optical depth retrieval outputs. Results for three scenarios under different atmospheric dust loading conditions are shown: on the left side, the χ2 values of the model-observation curve fitting in the $r_{eff} - \tau_0$ parameter space are mapped. The location of the minimum χ2 and the contours for the 68.3%, 95.4% and 99.7% confidence interval limits are indicated. On the right side, the Navcam retrieved sky radiances (gray) and the best fitting model curve (red) are graphed, together with the binned observation data (black) and the error-bars representing the absolute calibration uncertainty associated to the imager (12%).*

## 4. Results and discussion

In this section we present the results obtained in this study on the resulting effective radius and dust optical depth (Table 4). The seasonal variation of these parameters along MY 31 to 33 is evaluated (Figure 5 and 6) and the outputs are put into context by comparing with previous studies from other authors. Finally, a discussion is provided regarding the sensibility of the model and the uncertainties involved in the retrieval procedure.

4.1 Aerosol optical depth
From the retrieved seasonal behaviour of the optical depth (Figure 5, bottom), it can be appreciated the gradual decrease corresponding to the low dust opacity season, when the optical depth shifts from initial values of about $\tau \sim 0.75$ for $L_s$ around 40º, down to its minimum ($\tau \sim 0.4$) at $L_s = 135$º. After this point, a noticeable increase is appreciated right before $L_s = 150$º to values of $\tau \sim 0.75$, with a maximum retrieved opacity of $\tau \sim 1.0$ at 165º of solar longitude. A second period of dust enhanced activity can be observed after $L_s = 200$º, where there is a steep increase in dust opacities with $\tau$ scaling from values close to 0.8 up to greater than 1.25, corresponding to the maximum optical depth within the season cycle. Atmospheric dust loading drops down back to $\tau = 0.8$ near $L_s \sim 300$º, before a third dust enhanced activity period can be observed around $L_s = 325$º, when there is a subtle increase to $\tau$ near 1.0; before a final descent at the end of the year (only data for MY31) down to 0.70.

This seasonal behaviour of the optical depth agrees with previous descriptions for long-term dust optical depth retrievals by different missions and instrumentation since MY 12 (Viking Lander 1 and 2) for the periods without global dust storms (see e.g., Figure 10.3 in Khare et al., 2017). In particular, for MSL mission, both interannual and seasonal values of $\tau$ derived with Navcam show an overall good agreement with other MSL instrument data-set results published by other authors e.g.: Mastcam (Lemmon, 2014), REMS UV photodiodes (Smith et al., 2016), and Chemcam (McConnochie et al., 2017).

Table 4. Results of the retrieval

| MSL SOL | Solar Longitude [deg] | Martian Year | Local True Solar Time | Effective radius, $r_{eff}$ [μm] | Dust Column Optical Depth, $\tau_0$ | Reduced $\chi^2$ |
|---|---|---|---|---|---|---|
| 21 | 162.0 | 31 | 14:53:49 | $1.48^{+0.30}_{-0.29}$ | $0.95^{+0.06}_{-0.18}$ | 0.35 |
| 24 | 163.6 | 31 | 13:57:47 | $1.48^{+0.19}_{-0.20}$ | $0.76 \pm 0.04$ | 0.57 |
| 29 | 166.4 | 31 | 13:51:24 | $1.28 \pm 0.18$ | $0.72 \pm 0.04$ | 0.10 |
| 39 | 172.0 | 31 | 14:06:02 | $1.24^{+0.19}_{-0.18}$ | $0.70 \pm 0.04$ | 0.33 |
| 41 | 173.1 | 31 | 14:20:13 | $1.38 \pm 0.20$ | $0.77 \pm 0.05$ | 0.14 |
| 43 | 174.2 | 31 | 14:54:12 | $1.26 \pm 0.19$ | $0.72 \pm 0.05$ | 0.40 |
| 48 | 177.1 | 31 | 15:07:57 | $1.10 \pm 0.23$ | $0.85^{+0.09}_{-0.10}$ | 0.52 |
| 52 | 179.4 | 31 | 14:50:16 | $1.38^{+0.19}_{-0.20}$ | $0.70 \pm 0.04$ | 0.43 |
| 57 | 182.3 | 31 | 15:43:47 | $1.34 \pm 0.21$ | $0.66 \pm 0.05$ | 0.13 |
| 102 | 209.5 | 31 | 16:00:30 | $2.02^{+0.14}_{-0.13}$ | $1.38^{+0.11}_{-0.10}$ | 0.55 |
| 122 | 222.2 | 31 | 14:57:39 | $1.38 \pm 0.21$ | $0.99 \pm 0.08$ | 0.60 |
| 147 | 238.2 | 31 | 13:26:20 | $1.48 \pm 0.21$ | $0.91 \pm 0.06$ | 0.34 |
| 166 | 250.6 | 31 | 13:45:56 | $1.30 \pm 0.20$ | $0.93 \pm 0.07$ | 0.50 |
| 324 | 346.5 | 31 | 14:03:12 | $1.26^{+0.19}_{-0.18}$ | $0.68 \pm 0.04$ | 0.41 |
| 333 | 351.3 | 31 | 15:25:13 | $1.14 \pm 0.19$ | $0.64^{+0.05}_{-0.04}$ | 0.54 |
| 340 | 355.0 | 31 | 14:18:26 | $1.26 \pm 0.19$ | $0.66 \pm 0.04$ | 0.19 |
| 344 | 357.0 | 31 | 13:39:20 | $1.18 \pm 0.19$ | $0.72 \pm 0.04$ | 0.49 |
| 349 | 359.7 | 31 | 15:53:35 | $1.38 \pm 0.18$ | $0.72 \pm 0.05$ | 1.01 |
| 358 | 4.2 | 32 | 14:13:42 | $1.36^{+0.19}_{-0.20}$ | $0.77 \pm 0.05$ | 0.52 |
| 369 | 9.7 | 32 | 14:26:51 | $1.42 \pm 0.20$ | $0.79 \pm 0.05$ | 0.26 |
| 372 | 11.2 | 32 | 13:11:46 | $1.52 \pm 0.21$ | $0.70 \pm 0.04$ | 0.28 |
| 383 | 16.6 | 32 | 15:34:26 | $1.34^{+0.21}_{-0.23}$ | $0.70^{+0.05}_{-0.07}$ | 0.60 |
| 390 | 20.0 | 32 | 14:49:40 | $1.34^{+0.19}_{-0.20}$ | $0.70 \pm 0.05$ | 0.31 |
| 406 | 27.7 | 32 | 13:15:33 | $1.30 \pm 0.19$ | $0.70 \pm 0.04$ | 0.36 |
| 412 | 30.5 | 32 | 14:39:55 | $1.24^{+0.19}_{-0.18}$ | $0.64 \pm 0.04$ | 0.37 |
| 419 | 33.8 | 32 | 15:22:30 | $1.38^{+0.22}_{-0.23}$ | $0.74^{+0.06}_{-0.07}$ | 0.35 |
| 426 | 37.1 | 32 | 15:43:53 | $1.22^{+0.40}_{-0.41}$ | $0.77^{+0.22}_{-0.21}$ | 0.16 |
| 433 | 40.3 | 32 | 13:22:53 | $1.30 \pm 0.19$ | $0.72 \pm 0.04$ | 0.39 |
| 440 | 43.5 | 32 | 14:05:36 | $1.18^{+0.19}_{-0.18}$ | $0.72 \pm 0.05$ | 0.25 |
| 454 | 49.9 | 32 | 13:37:08 | $1.04 \pm 0.14$ | $0.58 \pm 0.03$ | 0.25 |

| MSL SOL | Solar Longitude [deg] | Martian Year | Local True Solar Time | Effective radius, $r_{eff}$ [μm] | Dust Column Optical Depth, $\tau_0$ | Reduced $\chi^2$ |
|---|---|---|---|---|---|---|
| 470 | 57.2 | 32 | 14:44:26 | $0.98 \pm 0.14$ | $0.50 \pm 0.03$ | 0.30 |
| 494 | 67.9 | 32 | 13:18:00 | $1.14 \pm 0.16$ | $0.50^{+0.02}_{-0.03}$ | 0.26 |
| 527 | 82.8 | 32 | 14:21:28 | $0.98 \pm 0.14$ | $0.50 \pm 0.03$ | 0.62 |
| 545 | 90.9 | 32 | 15:10:31 | $1.04 \pm 0.17$ | $0.48 \pm 0.03$ | 0.58 |
| 552 | 94.1 | 32 | 13:48:00 | $0.88 \pm 0.14$ | $0.49 \pm 0.03$ | 0.77 |
| 563 | 99.1 | 32 | 14:34:31 | $0.90 \pm 0.12$ | $0.45 \pm 0.02$ | 0.45 |
| 569 | 101.9 | 32 | 14:20:15 | $0.92 \pm 0.14$ | $0.43 \pm 0.02$ | 0.59 |
| 589 | 111.2 | 32 | 13:22:50 | $1.00^{+0.13}_{-0.14}$ | $0.45 \pm 0.02$ | 0.36 |
| 631 | 131.4 | 32 | 13:50:16 | $0.86 \pm 0.12$ | $0.41 \pm 0.02$ | 0.44 |
| 637 | 134.4 | 32 | 13:41:18 | $0.88 \pm 0.14$ | $0.41 \pm 0.02$ | 0.67 |
| 647 | 139.5 | 32 | 13:43:25 | $1.08 \pm 0.16$ | $0.43 \pm 0.02$ | 0.52 |
| 657 | 144.6 | 32 | 14:07:46 | $1.10 \pm 0.16$ | $0.62 \pm 0.03$ | 0.32 |
| 662 | 147.2 | 32 | 14:49:43 | $0.92 \pm 0.15$ | $0.64 \pm 0.04$ | 0.40 |
| 668 | 150.3 | 32 | 14:29:06 | $0.92 \pm 0.16$ | $0.64 \pm 0.04$ | 0.78 |
| 672 | 152.4 | 32 | 13:56:42 | $1.18 \pm 0.19$ | $0.66 \pm 0.04$ | 0.59 |
| 685 | 159.4 | 32 | 15:04:35 | $1.10 \pm 0.18$ | $0.66 \pm 0.05$ | 0.30 |
| 705 | 170.5 | 32 | 14:50:54 | $1.42^{+0.30}_{-0.29}$ | $0.99^{+0.11}_{-0.15}$ | 0.40 |
| 733 | 186.7 | 32 | 15:27:01 | $1.32 \pm 0.22$ | $0.75 \pm 0.06$ | 0.37 |
| 747 | 195.0 | 32 | 14:44:14 | $1.20 \pm 0.19$ | $0.77 \pm 0.05$ | 0.57 |
| 864 | 269.8 | 32 | 13:48:01 | $1.26 \pm 0.21$ | $0.91 \pm 0.07$ | 0.39 |
| 952 | 323.9 | 32 | 13:35:01 | $1.32 \pm 0.19$ | $0.77^{+0.04}_{-0.05}$ | 0.34 |
| 964 | 330.8 | 32 | 14:52:45 | $1.34 \pm 0.34$ | $0.95^{+0.08}_{-0.23}$ | 0.40 |
| 984 | 341.9 | 32 | 13:21:25 | $1.34^{+0.20}_{-0.21}$ | $0.93 \pm 0.06$ | 0.32 |
| 1067 | 24.1 | 33 | 13:58:10 | $1.24 \pm 0.20$ | $0.75 \pm 0.05$ | 0.49 |
| 1104 | 41.4 | 33 | 13:51:37 | $1.18 \pm 0.17$ | $0.68 \pm 0.04$ | 0.22 |
| 1167 | 69.9 | 33 | 14:32:05 | $0.98 \pm 0.14$ | $0.52 \pm 0.03$ | 0.55 |
| 1262 | 113.3 | 33 | 15:45:07 | $1.06 \pm 0.19$ | $0.50 \pm 0.04$ | 0.65 |
| 1301 | 132.2 | 33 | 14:41:32 | $0.76^{+0.12}_{-0.11}$ | $0.47^{+0.02}_{-0.03}$ | 0.68 |
| 1376 | 171.9 | 33 | 15:08:58 | $1.14^{+0.20}_{-0.19}$ | $0.77 \pm 0.06$ | 0.41 |
| 1433 | 205.6 | 33 | 13:42:01 | $1.14 \pm 0.19$ | $0.89 \pm 0.06$ | 0.42 |
| 1468 | 227.8 | 33 | 14:47:57 | $1.52^{+0.35}_{-0.34}$ | $1.28 \pm 0.20$ | 0.49 |
| 1503 | 250.5 | 33 | 13:33:52 | $1.08 \pm 0.19$ | $0.93 \pm 0.07$ | 0.44 |
| 1571 | 294.1 | 33 | 12:59:09 | $0.96 \pm 0.13$ | $0.80^{+0.05}_{-0.04}$ | 0.23 |
| 1604 | 314.2 | 33 | 13:19:28 | $0.94 \pm 0.15$ | $0.91 \pm 0.06$ | 0.23 |
| 1646 | 338.3 | 33 | 13:39:02 | $0.98 \pm 0.17$ | $0.97^{+0.07}_{-0.08}$ | 0.30 |

*Note: MSL mission sol number, solar longitude (in degrees), Martian Year (following Clancy et al. 2000 convention), local true solar time, particle size distribution effective radius (in microns), dust aerosol column optical depth at surface (referenced to 880nm), and reduced $\chi^2$ value, $\chi^2_\nu$. The uncertainties of the effective radius and optical depth were calculated for a 68.3% confidence limit for a $\chi^2$ distribution probability density function for each parameter.*

4.2 Dust effective radius

On Figure 6 the interannual and seasonal variation of dust aerosol particle size distribution effective radius are shown. These results are put into context by comparing with other MSL retrievals: Chemcam passive sky spectral observations (McConnochie et al., 2017), and REMS UV photodiodes (Vicente-Retortillo et al., 2017).

Regarding the seasonal behaviour (Figure 6, bottom) of the effective radius, during the aphelion season ($L_s$ = 0º to 180º) it presents first a steady decrease with $r_{eff}$ varying from 1.40 µm to reaching minimum values of 0.80 to 0.90 µm at $L_s \sim 130$º; followed by a steep increase to $r_{eff}$ of 1.50 µm at approximately $L_s$ = 180º; which also corresponds to an enhancement in the dust column opacity. After this, a slight decrease down to effective radii of about 1.20 µm can be identified until $L_s \sim 200$º; before larger particle sizes of $r_{eff} \sim 1.50$ µm (especially in MY31) are observed in the proximity of $L_s$ = 230º. Following this period, a new drop can be appreciated with radius values falling down to 1.0 µm at $L_s \sim 300$º (only MY33 data), before a final increase to radii of 1.30 µm at $L_s$ = 350º. A discrepancy in this seasonal behaviour with respect to Chemcam and REMS UV results can be appreciated, especially for the second half of the year. A possible explanation for this may be found in the analysis of the interannual variation (Figure 6, top). The lack of previous data available for comparison for the perihelion season ($L_s$ = 180º to 360º) of MY33 (sol > 1400) does not allow us to evaluate the actual level of discrepancy; however, results corresponding to the first half of that year (sols between 1000 and 1400) showed that the retrieved effective radii were smaller comparing to the same period of the previous year, and therefore it might reduce this deviation from those studies.

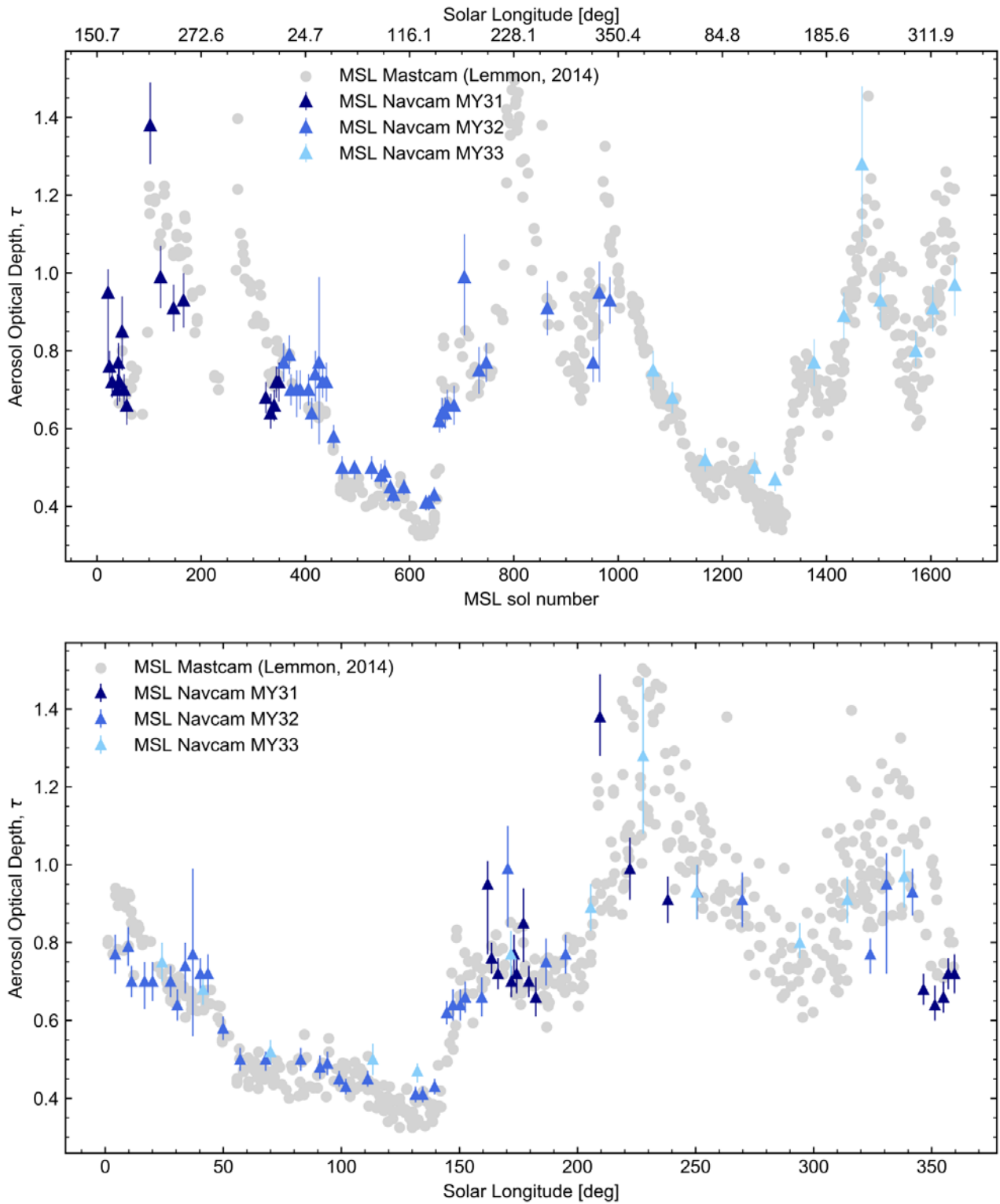

*Figure 5. Dust aerosol column optical depth derived with MSL Navcam. Results for the interannual (top) and seasonal (bottom) variations of the dust column optical depth obtained from the 65 Navcam observations, covering a period of almost 3 Martian Years, from sol 21 ($L_s$=162º, MY31) to sol 1646 ($L_s$=338º, MY33), are presented in these graphs. The results are compared to optical depth retrievals from MSL Mastcam using direct Sun Imaging for the same period (Lemmon, 2014). For comparison purposes with Mastcam, the column optical depths retrieved in our study are referenced to a wavelength of 880 nm.*

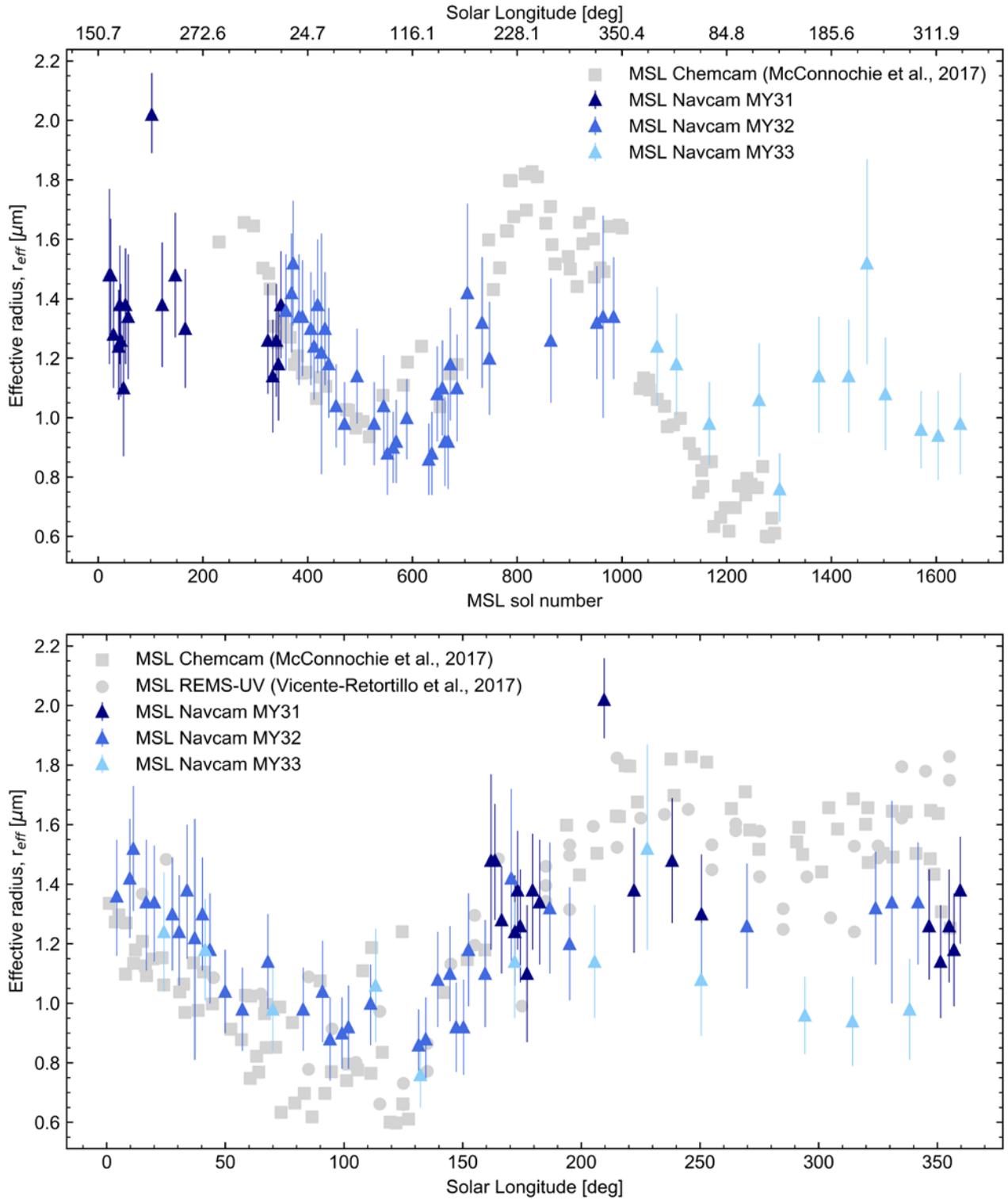

*Figure 6.* Variation of the dust effective radius. The interannual (top) and seasonal (bottom) behaviour for the effective radius of dust aerosol particle size distribution obtained in this work are shown in these figures for Martian Years 31, 32 and 33. The results are compared with previous retrievals with MSL Chemcam (squares) and MSL REMS UV (circles).

4.3 Correlation between particle effective radius and aerosol optical depth

Results of the retrieval for dust column optical depth and aerosol particle size distribution effective radius are shown in Figure 7. The coefficient of determination ($R^2$) was calculated for these two variables and a value of 0.49 was obtained, which indicates a low to medium correlation.

When comparing to previous studies for MSL mission, the $R^2$ values for Chemcam and REMS UV instruments retrievals were of 0.69 and 0.67, respectively (McConnochie et al., 2017; Vicente-Retortillo et al., 2017). However, it is worth mentioning that the sol period covered by these outputs were different, being the latest available data for Chemcam corresponding to sol 1291, and sol 1159 for REMS UV. If the results for the observation set used in this study (latest sol is 1646) are limited to those dates, correlation coefficients of 0.63 and 0.69 are obtained for each ending sol period, respectively.

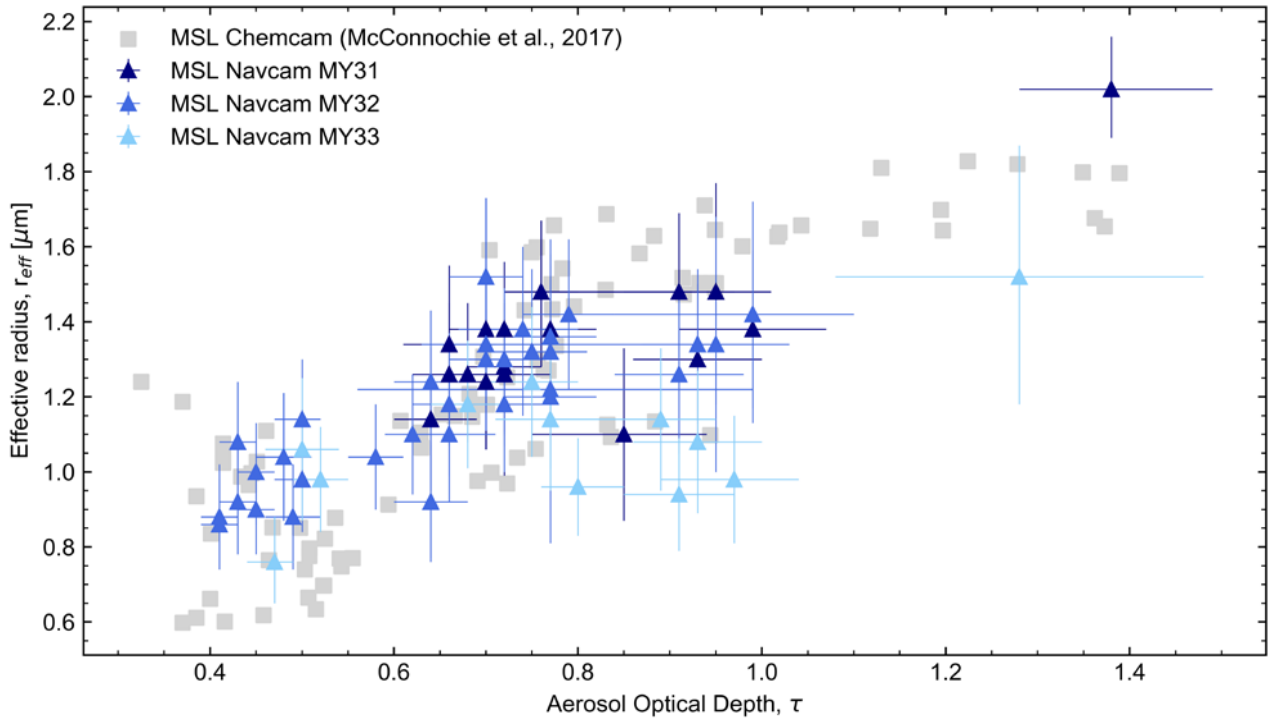

*Figure 7. Relationship between dust particle effective radius and aerosol optical depth. The dust particle effective radius is represented as a function of the derived aerosol column optical depth. For comparison purposes, the retrievals from MSL Chemcam are also included in this figure.*

4.4 Sensitivity study

In the retrieval procedure described above, some assumptions were made on part of the input parameters required by the sky radiance curve modelling. In the following paragraphs, the robustness of the derived results is evaluated by studying the sensitivity of the outputs to variations of these parameters.

*Parameter retrieval*. From the best-fitting regions of the $\chi^2$ maps for the $r_{eff} - \tau_0$ parameter space presented on Figure 4, it can be appreciated that the retrieval procedure presents more sensitivity to the column optical depth than to the effective radius. This is due to the different influence that each free parameter has on the modelled curves: while the column optical depth input defines the overall values of radiance factor *I/F* of the sky brightness function, the effective radius parameter mainly controls the curvature of the curve for the evaluated scattering angle range. In order to estimate the effect of the $\tau_0$ parameter on the $r_{eff}$ outputs, a simulation was performed in which column optical depth inputs were set in accordance with the values derived by Mastcam direct Sun-imaging for the nearest sol (Lemmon, 2014). Regarding the dust column optical depths, the average difference between the retrievals of the base simulation (2 free parameters) and Mastcam $\tau$ records were around 10%; while for the dust particle effective radius the mean difference between these retrievals were less than 16%, being the highest discrepancy values (30% to 45%) mainly located in the $L_s$ = 120º-160º and 320º-330º windows.

*Sky radiance sampling path.* From the different sky radiance sampling paths presented on Figure 3, we selected the diagonal direction in an analogous way to Soderblom et al., (2008a, 2008b). In order to estimate the sensitivity of the results to the chosen sampling direction, we compared the outputs of our base model to simulation results from a retrieval following the principal plane direction (sky points with azimuth angle equal to Sun's azimuth). This comparison returned an average difference of 11% for the column optical depth and 8% for the effective radius parameter. The seasonal variation of this discrepancy was evaluated and showed no particular relationship for the $\tau_0$ parameter, whereas for the effective radius the highest discrepancy percentages (15% to 20%) where concentrated within the $L_s$ 50º to 140º period (dust low opacity season). This is due to the brightness distribution around the solar disc aureole during this period, when high intensities are mainly located at very low scattering angles, as it can be appreciated on Figure 1; and it is precisely within this part of the curve (scattering angle $\theta < 10º$) where the effective radius parameter has its main influence in the sky brightness function. Regarding the goodness of fit reduced $\chi^2$ parameter, the principal plane sampling direction values were in average 5% lower than the diagonal ones.

*Aerosol particle shape.* Regarding the selected shape of the dust aerosol particle model, previous studies showed that the shape of the particle has negligible influence at the forward scattering region (scattering angles up to 30º) (e.g., Hansen and Travis, 1974; Pollack et al.; 1995). The sensitivity of the outputs to the selected particle shape was evaluated by comparing results of two simulations using spherical and cylindrical particles with diameter to length aspect ratio of 1.0. This comparison returned very similar seasonal patterns for both dust opacity and effective radius parameters and the average differences with respect to the best fitting values were of less than 7% for column optical depth and 13% for the effective radius; both quantities were contained within the uncertainty region of the nominal scenario. The average difference in the reduced $\chi^2$ values were approximately less than 2% lower for the spherical particle simulation than the base scenario (cylindrical).

*Effective variance of the aerosol particle size distribution.* For the sensitivity study of the results to changes in the effective variance ($v_{eff}$) of the particle size distribution, we performed additional retrievals with the aerosol model set to $v_{eff}$ equal 0.4 and 0.5 (e.g., Tomasko et al., 1999). The outputs of these simulations were compared to the nominal model ($v_{eff} = 0.3$); for the $v_{eff} = 0.4$ simulation, an average difference of less than 2% for the dust column opacity best retrieval and around 11% for the effective radius was obtained, while for the $v_{eff} = 0.5$ run mean variations of about 3% and 13% for $\tau_0$ and $r_{eff}$ were appreciated, respectively. The resulting differences derived from these simulations were all located within the uncertainty range of the outputs in the nominal scenario. For the reduced $\chi^2$ parameter, the obtained average values in the $v_{eff} = 0.3$ model were 4% and 7% greater than the 0.4 and 0.5 effective variance models, respectively.

*Vertical distribution of the aerosol optical depth.* This is directly related to the aerosol vertical distribution (dust mass mixing ratio) (Heavens et al., 2011), which was modelled using the modified *Conrath* profile in [5]. This depended on the total column optical depth at surface, $\tau_0$, and the $l$ and $v$ constants, which controlled the dust layer maximum altitude and the vertical profile shape. Several simulations were performed for limit values of these parameters: for dust layer top altitudes of 40 km ($l = 1.75$) and 80 km ($l = 0.875$), and for vertical profiles with exponential ($v = 0.1$) and step ($v = 0.001$) shapes. The outputs of these simulations showed that the model had no sensitivity to such changes.

*Surface albedo.* In the radiative transfer model used in this study the surface albedo parameter was set to an average value of 0.20 for the Gale Crater region. For surface based upward-looking observations, it can be expected that surface reflectivity would have little impact on the retrieved image intensity (Pollack et al., 1995); in contrast with downward-pointing observations made from orbit, on which the reflection properties of the ground need to be separated from atmospheric dust aerosol scattering phase function (Tomasko et al., 1999). *S*everal retrievals were performed for different surface albedo values ranging from 0.10 to 0.50, covering the possible values for Gale Crater (Anderson and Bell, 2010). The results of these retrievals showed that both the effective radius and dust column opacity had no sensitivity to such variations.

# 5. Conclusion

We have shown in this paper that the Navigation Cameras onboard MSL rover Curiosity can be used to estimate the atmospheric dust opacity and constrain the aerosol particle size effective radius. For this study, a total of 65 Sun pointing observations were selected, spanning from sol 21 to sol 1646 and covering 2.5 Martian Years.

Radiometric calibration and geometric reduction were performed on the MSL Navcam images following the in-flight calibration process derived for MER Navcam as they are build-to-print copies. The calibration outcome image data were validated against MSL Mastcam observations. The observed sky brightness as a function of the scattering angle data were compared against modelled curves from a multiple scattering radiative transfer model of Mars' atmosphere, in order to retrieve the optical depth and aerosol effective radius parameters generating the best fitting curve.

The seasonal behaviour of the dust column opacity and particle size distribution effective radius were obtained and evaluated. Significant seasonal variations were detected from these retrievals and a positive correlation between high optical depth values with larger particle size was inferred. The results of this work were compared with previous studies using different instrumentation on-board the MSL rover and presented an overall good agreement.

We can take advantage of the observational versatility of the engineering cameras, their capability of covering wide sceneries and their frequent nominal use rate, in order to provide atmospheric studies with large and varied sets of sky observations that can contribute in the characterisation, modelling and better understanding of the airborne dust and its role in Mars' atmosphere.

Ongoing research with further MSL engineering cameras data (using both Navcam and Hazcam) include the use of the 360º sky survey observations to study the sky radiance at high scattering angles (up to approximately 160º), in combination with other MSL imagers in order to characterise the shape of the dust aerosol particles.

**Acknowledgements**
This work was supported by the Spanish project AYA2015-65041-P (MINECO/FEDER, UE), Grupos Gobierno Vasco IT-765-13, and Diputación Foral de Bizkaia - Aula EspaZio Gela. We wish to thank Professor Mark T. Lemmon for providing the MSL Mastcam optical depth values shown in this work for comparison.